# REPRESENTATION OF QUDITS ON A RIEMANN SPHERE

Rahul Bijurkar*

**ABSTRACT –**

In quantum computation and information science, the geometrical representations based on the Bloch sphere representation for transformations of two state systems have been traditionally used. While this representation is very useful for the two state qubit, it cannot be generalized easily to multiple states like that of qudits and when it is generalized, it looses its simple geometrical representation.
    This paper proposes the use of an alternative representation in quantum information and computation for qubits as well as qudits based on the Majorana representation on the Riemann Sphere, which preserves the simple sphere representation.

-------------------------------------------

## 1. Introduction

In recent years there has been considerable interest in the theory of quantum computation and information. It has not only shown a way of developing faster algorithms for solving different problems but also has contributed to bringing about a fundamental shift in the way theory of computation is perceived. At the same time, ideas developed in quantum information theory are helping to gain new understanding of the fundamental tenets of quantum physics.

The quantum computer, like its classical counterpart, is based on two level systems represented as 0 and 1. A classical bit is the most basic unit of information and can have one of the two values 0 or 1. A quantum bit or qubit is similarly the most basic unit of quantum information but a qubit can exist as a superposition of the two possible states, represented as,

$$|\psi\rangle = a|0\rangle + b|1\rangle \text{ with } |a|^2 + |b|^2 = 1 \quad \text{-----(1)}$$

The qubit can be represented geometrically as a point on a Bloch sphere by specifying the three polar coordinates [3,4] and up to a global phase factor can be written as

$$|\psi\rangle = \cos \phi/2 \, |0\rangle + e^{-i\theta} \sin \phi/2 \, |1\rangle \quad \text{-----(2)}$$

__________________

* The author is an Electrical Engineer with Bhilai Steel Plant, Steel Authority of India Ltd., India

Any transformation on this qubit can then be represented by appropriate rotations of the Bloch sphere. This representation is very useful and convenient in case of a single qubit and gives a clear geometrical picture of a qubit and expresses quantum states in terms of the expectation values of observables. But the Bloch sphere representation cannot be generalized easily to more than one qubit or multi-valued qudits. While some work has been done in generalizing the Bloch representation for N-level systems, it is seen that the simple sphere representation no longer applies and asymmetric structures appear in N-level systems which do not have rotational invariance [8,9].

Another representation of a two level system or qubit can be given in terms of points on a Riemann sphere [4,7]. The Riemann sphere is a complex projective space, which is formed by stereographically projecting the complex plane on to the sphere and including the point at infinity. For one qubit, this representation is very similar to the Bloch sphere representation and these representations can be transformed into each other [4]. A qubit can be represented as a point on Riemann sphere and a qubit transformation is obtained by the set of rotations of the sphere. These transformations are the Mobius transformations and belong to SU(2) group.

## 2. Representing a Qubit on Riemann sphere

Any complex number can be represented by a point in the complex (Argand) plane. A qubit represented by eq-(1) is defined by its two coefficients a and b which are complex valued in general. A quantum state is generally not changed if it is multiplied by a scalar. Hence, it is actually the ratio a/b that defines any qubit state $\psi$. This can be represented geometrically as a point α = a/b, in the extended complex plane $C \cup \infty$. This plane can be stereographically projected onto a sphere.

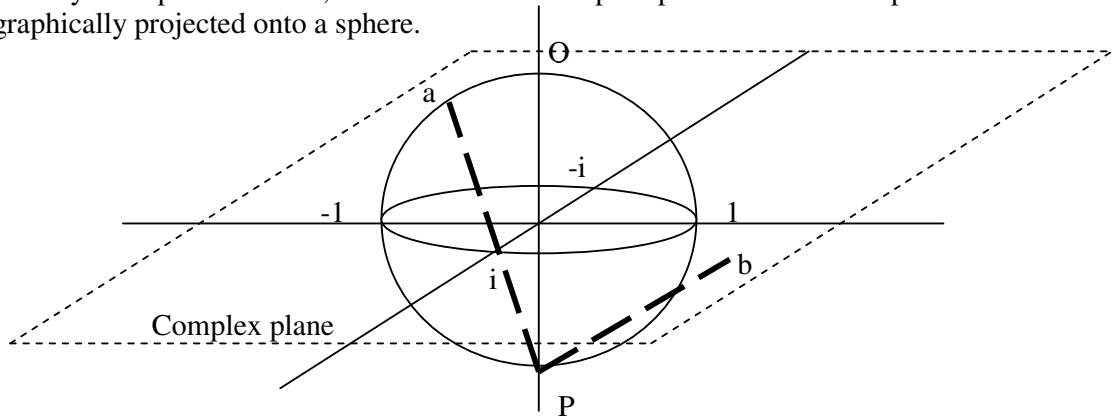

Fig-1: Stereographic projection of $C \cup \infty$ on a sphere.

As shown in Fig.–1, if a complex plane is taken as the equatorial plane of a unit sphere and each point of the complex plane is projected from the point P onto the sphere, it is seen that the origin projects to the point Q, all the points inside the unit circle project onto the upper hemisphere and all the points outside the unit circle project to lower hemisphere. The point at infinity projects to the point P. Hence it is seen that the extended complex plane $C \cup \infty$ can be represented as points on the sphere. This Sphere is called the Riemann sphere [7]. A qubit is thus represented by a point on the Riemann sphere.

## 3. Mobius transformations as transformations on a qubit

The Mobius transformation is a bijective conformal mapping of the extended complex plane or the Riemann sphere. These transformations map the Riemann sphere to itself, preserving the angles and orientation. A Mobius transformation thus transforms a point on the Riemann sphere and is given by

$$z \mapsto \frac{az+b}{cz+d} \qquad \text{-----(3)}$$

where a,b,c&d are any complex numbers satisfying $ad - bc \neq 0$.
This transformation can be expressed in the matrix form as

$$\Phi = \begin{pmatrix} a & b \\ c & d \end{pmatrix} \qquad \text{-----(4)}$$

with the condition that the matrix is non singular ($ad - bc \neq 0$).

The Mobius transformations form a group under association called the Mobius group. This group is isomorphic to the projective General Linear group and under the restriction of $\det(\Phi) = 1$, the group is isomorphic to the projective SL(2,C) group. The group SL(2,C) forms the universal cover of the Mobius group. A special case of this group is the SU(2) group formed when the matrices taken to be unitary. Hence the SU(2) group of transformations form a subgroup of the Mobius group called the special unitary Mobius transformations.

When the matrix $\Phi$ is considered to be unitary with the additional constraint $\det(\Phi) = 1$, the matrix takes the general form,

$$\Phi = \begin{pmatrix} a & b \\ -b^* & a^* \end{pmatrix} \text{ with aa* + bb* = 1.} \qquad \text{-----(5)}$$

Thus the super unitary Mobius transformation can be expressed as a special case of a rotation of the Riemann sphere given by,

$$z \mapsto \frac{az+b}{-b^*z+a^*} \qquad \text{----- (6)}$$

Hence any qubit represented by a point on Riemann sphere can be transformed by the Mobius transformations and give the set of unitary transformations which define the various gates in quantum computation [4].

## 4. Majorana Representation of spin states

Ettore Majorana published a fascinating paper in 1932 [6,7] showing the relation between the quantum mechanics of large spin systems like atoms and polynomials. He used the sphere representation to aid the calculation of the transition probabilities for atoms. He showed in relation to atomic spins that the states of an atom with spin n/2 can be represented by a set of n spin – ½ particles. This is geometrically represented by n points on the Riemann sphere. This representation has the advantage that it no longer needs any privileged axis.

Using the orthonormal basis states of a spin – n/2 atom, the general state can be written as superposition of the basis states.

$$|\psi\rangle = a_0|0\rangle + a_1|1\rangle + a_2|2\rangle + \ldots\ldots a_{n-1}|n-1\rangle + a_n|n\rangle$$

$$= \sum_{\mu=0}^{n} a_\mu |\mu\rangle \qquad \text{----- (7)}$$

In the Majorana representation [5], the basis states of the (n+1) dimensional spin space is represented as terms of a monomial in a determinate z, where z can be complex valued. If the arbitrary $|\mu\rangle$ state is written as

$$|\mu\rangle \mapsto (-1)^{\mu} \sqrt{\left(\frac{n!}{(n-\mu)!\mu!}\right)} z^{\mu} \quad \text{----- (8)}$$

then,

$$p(\psi) = \sum_{\mu=0}^{n} a_{\mu} (-1)^{\mu} \sqrt{\left(\frac{n!}{(n-\mu)!\mu!}\right)} z^{\mu} \quad \text{----- (9)}$$

This Majorana polynomial in z has n complex roots and can be factorised as

$$p(\psi) = a_n \sum_{d=1}^{n} (z - \alpha_d) \quad \text{----- (10)}$$

The roots $\alpha_n$ of the polynomial $p(\psi) = 0$ are invariant under multiplication by any complex number and hence are invariant up to an overall phase and amplitude.

These n roots can be represented as n points on the Riemann sphere with each vector joining the origin to the point on sphere representing on spin ½ particle. Thus the spin n/2 particle is represented as n spin ½ particles like in the case of representation of spin 1/2 particle on Riemann sphere, any transformation on this state is given by the rotation of the sphere or Mobius transformations.

Hence it is seen that the Majorana representation is a complete visual representation of the wave function and therefore carries complete quantum information about the state of the system.

### 5. Multivalued Logics and Qudits

A qudit is defined as a d-level unit of information. In the same way that a qubit is used to define a two valued logic system, a qudit is used to define multi-valued logic systems. Such a system of qudits and multi-valued gates provide greater flexibility in the information storage and also reduces the time taken to implement a logic gate sequence. But this is at the cost of increased complexity of gate and finer control requirements of the levels. These ideas have been extensively studied [1,2] for their advantages and also because they generalize the qubit based computation system.

But in geometrically representing such qubits, the Bloch sphere representation, widely used for the qubit, cannot easily be extended to represent the qudits. At the same time it is seen that a qudit can be physically simulated in a straight forward way by a d- level atom with spin n/2 where n=(d-1).

This fact can be effectively used to extend the qubit representation on Riemann sphere to represent qudits on the Riemann sphere using the Majorana representation.

### 6. Qudit Representation on Riemann Sphere

The Majorana representation of a spin-n/2 particle can be utilized in a straight forward way to represent a d-level qudit., where d=n+1. In the Majorana representation, a qudit can be represented by the (d-1) – zeros of the Majorana polynomial. These are represented by the (d-1) points on the Riemann sphere.

A d-level qudit state can be written as –

$$|\psi\rangle = a_0|0\rangle + a_1|1\rangle + a_2|2\rangle + \ldots\ldots a_{d-1}|d-1\rangle$$

which can be represented as a point in the d-dimensional Hilbert space. On applying the Majorana transformation given by eq.-(8) on the above state, we get the Majorana polynomial,

$$p(\psi) = \sum_{\mu=0}^{n} a_\mu (-1)^\mu \sqrt{\left(\frac{n!}{(n-\mu)!\mu!}\right)} z^\mu \quad \text{where } n = (d-1) \quad \text{--(11)}$$

This polynomial has (d-1) zeros given by

$$p(\psi) = a_{(d-1)} \sum_{\mu=1}^{d-1} (z - \alpha_\mu) \quad \text{-----(12)}$$

These zeros are represented by (d-1) Majorana points on the Riemann sphere. It is therefore seen that the Majorana transformation transforms a point in the d-dimensional Hilbert space to (d-1) points on the Riemann Sphere. Thus a unitary transformation of a point on d-dimensional Hilbert space is equivalent to a unitary transformation of (d-1) Majorana points on the Riemann sphere. Any unitary transformation on the state $|\psi\rangle$ of a qudit can therefore be given by the Mobius transformation of the Majorana polynomial representing that state,

$$a_{(d-1)} \sum_{\mu=1}^{d-1} \xi_\mu z^\mu \mapsto a_{(d-1)} \sum_{\mu=1}^{d-1} \xi_\mu \left(\frac{az+b}{-b*z+a*}\right)^\mu$$

$$\mapsto a_{(d-1)} \sum_{\mu=1}^{d-1} \beta_\mu z^\mu \quad \text{----- (13)}$$

As specific examples, a three level qutrit is represented by a quadratic Majorana polynomial having two zeros represented by two points on the Riemann sphere. It is seen that a qutrit can be simulated by a spin-1 particle.

$$|\psi\rangle = a_0|0\rangle + a_1|1\rangle + a_2|2\rangle$$
$$P(\psi) = a_2 z^2 + a_1 z + a_0$$
$$= (z-\alpha_1)(z-\alpha_2)$$

The Mobius transformation f(z) = 1/z transforms the polynomial to
$$P^T(\psi) = a_2 + a_1 z + a_0 z^2$$

And hence simulates a NOT gate for a qutrit, giving a transformation similar to the matrix,

$$\text{NOT} \rightarrow \begin{pmatrix} 0 & 0 & 1 \\ 0 & 1 & 0 \\ 1 & 0 & 0 \end{pmatrix}$$

Similarly the Mobius transformation, $f(z) = \frac{z-1}{z+1}$ gives the Hadamard transform.

It is thus seen that any Mobius transformation can provide a gate for qudit computation. It is interesting to note that as the number of levels –d in a qudit increase, the dimensions of the unitary matrix required for transformation increases, ie, for d- level system, a d x d matrix is required for the unitary transformations. But in this representation, a simple complex function f(z) is needed for rotation of Riemann sphere requiring only 2 – parameters as per eq-(6). Also this transformation is same for a qubit, qutrit or any level of qudit.

## 7. Generalising to Multiple Qudits

When dealing with multiple qubits or qudits, similar representation may be used but the physical picture does not become clear. The n-qubits have $2^n$ states and can be represented by $2^n -1$ points or vectors on the Riemann Sphere. As in the case of Majorana representation for atoms, where spin-n/2 is represented by n spin ½ particles, the n 2-state qubits cannot be represented by n particles but have to be represented by $(2^n -1)$ particles. While representations of a single qudit are easily pictured in this representation, its generalization to multiple qudits may require higher dimensional complex projective spaces. The Riemann sphere is a Complex Projective space of dimension – 1 ie, $CP^1$. For n qudits, the space $CP^n$ may be needed to completely represent its $d^n$ states. Further work in this area may lead us to a general representation of multiple qudits. This work stresses on the possibility that the richness of the conformal theory can be used to better understand the ideas of quantum computation and information theory.

## 8. Conclusion

Geometrical representations of mathematical equations and transformation greatly help in simplifying the procedures as well as lead to better understanding of the processes going on. In quantum physics, and consequently in quantum computation and information science, the Bloch sphere representation for transformations of two state systems has been traditionally used. While this representation is very useful for two state systems, it cannot be generalized to multiple states and when it is generalized, it looses its simple geometrical representation.

This paper proposes the use of an alternative representation in quantum information and computation based on the Majorana representation on the Riemann Sphere. While this representation is the same as Bloch sphere representation for two state systems, the advantage of this representation is that when generalized to multiple states, it does not loose its simple geometrical structure. This is shown in the paper by generalizing the qubit representation to a qudit.